\documentclass[aps,prl,floats,twocolumn,floatfix,showpacs]{revtex4}
\usepackage{epsf,graphicx}
\usepackage{amssymb}
\usepackage{amsmath}
\usepackage{eepic}

\newlength{\piclen}
\begin{document}

\title{Sr$_2$(Ba$_2$)VO$_4$ under pressure -- an orbital 
switch and potential $d^1$ superconductor}

\author{R.\ Arita$^a$, A.\ Yamasaki$^b$,  K.\ Held$^b$,  J. \ Matsuno$^a$,
K.\ Kuroki$^c$}
\affiliation{$^a$RIKEN, Wako, Saitama 351-0198, Japan\\
$^b$ Max-Planck-Institut f\"ur
Festk\"orperforschung, 70569 Stuttgart, Germany\\
$^c$ University of Electro-Communications
1-5-1 Chofugaoka, Chofu-shi Tokyo 182-8585, Japan}

 \date{\today}

\begin{abstract}
We study  Sr$_2$(Ba$_2$)VO$_4$   under high pressure 
by means of the local density approximation + dynamical 
mean field theory method. While Sr$_2$VO$_4$ is a 1/6-filling three-band 
system at ambient pressure with a small level splitting between 
the $d_{xy}$- and $d_{yz/zx}$-bands, we show that an orbital 
polarization occurs under uniaxial pressure, resulting
in dramatic changes of the magnetic, optical, and transport properties.
When pressure is applied in the $c$-direction, a $d^1$ analog of $d^9$ 
cuprates is realized, making Sr$_2$(Ba$_2$)VO$_4$  a  possible candidate
for a $d^1$ superconductor. Experimentally, this uniaxial pressure can be 
realized by growing Ba$_2$VO$_4$ on a substrate with lattice constant 
4.1-4.2 \AA.
\end{abstract}


\maketitle

Since the discovery of high temperature superconductivity
in cuprates\cite{Bednorz86}, strongly correlated electron systems  (SCES) and
their intriguing magnetic, dielectric, optical 
and transport properties 
have been at the center of solid state research.
Hence, a quantitative reliable
calculation of correlation effects from first-principles 
is one of the most important challenges.
This is particular difficult since 
the standard local density approximation (LDA) in the 
framework of density functional theory\cite{Kohn}
fails if electronic correlations are strong.
Recently however, a variety of attempts which go beyond LDA have been 
undertaken, and many successes have been achieved\cite{beyondLDA}. 

The next step in this direction is the
 (theoretical) materials design of SCES
with specific properties
and the simulations of SCES under extreme conditions.
As a  touchstone for such attempts, 
we study the electronic 
structure of Sr$_2$VO$_4$ and  Ba$_2$VO$_4$  under high pressure by means of 
the LDA+DMFT (dynamical mean field theory) method\cite{Anisimov}, one
of the most widely used approaches for
realistic calculations of SCES \cite{Held}. 

The reason why we focus on Sr$_2$VO$_4$ is twofold.
First, Sr$_2$VO$_4$ is a layered perovskite,
as  cuprates and ruthenates\cite{Maeno03}
which show unconventional superconductivity.
Second, the challenge to synthesize
single-crystalline Sr$_2$VO$_4$
has been overcome quite recently: One of the authors
and his coworkers\cite{Matsuno05} employed epitaxial growth 
techniques for growing a thin Sr$_2$VO$_4$ film. 
Hence, a detailed investigation of 
the electronic structure becomes now possible.

As for the first point, in fact, Sr$_2$VO$_4$ 
attracted attention because it is a "dual" material of 
La$_2$CuO$_4$. Namely, the former has one 3$d$ electron 
per V site ($d^1$ system), while the latter has nine
3$d$ electrons per Cu site ($d^9$ system). 
However, as was already pointed out by Pickett {\it et al.} 
in 1989\cite{Pickett}, there is a big difference between 
these oxides. While La$_2$CuO$_4$ is a 1/2-filling 
single-band ($d_{x^2-y^2}$) system, Sr$_2$VO$_4$ is 
a 1/6-filling three-band ($d_{xy/yz/zx}$)
system, since the level splitting of $t_{2g}$ in
the latter material is 
much smaller than that of $e_g$ in the former.
Therefore, the theoretical idea of unconventional
superconductivity in  Sr$_2$VO$_4$ has been dismissed.

Also experimentally, Sr$_2$VO$_4$ and La$_2$CuO$_4$  behave indeed
differently. Magnetic properties and transport 
properties were measured for a polycrystal of Sr$_2$VO$_4$ in the
early 1990's\cite{Nozaki}, indicating that Sr$_2$VO$_4$ is 
an antiferromagnetic insulator (semiconductor) with a
low N\'eel temperature $T_N\sim$ 45K. But in contrast to La$_2$CuO$_4$, a 
small ferromagnetic moment 
was also observed. In recent measurement of the optical conductivity 
for a single-crystalline thin film, 
a small gap structure, i.e., a peak around 1 eV and
a shoulder around 0.5 eV was observed\cite{Matsuno05}.

Theoretically,  several first-principles 
calculations  beyond LDA were performed,
 and it was confirmed that the 
relation between Sr$_2$VO$_4$ and La$_2$CuO$_4$ is not
dual\cite{Imai05,Weng05}.
Especially, Imai {\it et al.} carried out an LDA+PIRG 
(path integral renormalization group) calculation and found 
a nontrivial orbital-stripe order\cite{Imai05}. 
These orbital degrees of freedom are irrelevant in cuprates.

The motivation of the present study is based on the
following idea: While Sr$_2$VO$_4$ 
and La$_2$CuO$_4$ are certainly not dual
at ambient pressure,  we
hope to change this by changing the atomic configration
of the octahedron around the V ion. Here we consider 
to apply high uniaxial pressure,
synthesize films on substrates with appropriate
lattice constant, and introduce chemical pressure
(substitute Sr by Ba).
If the level splitting between the $d_{xy}$ and $d_{yz/zx}$ orbitals
becomes larger, it will lead to an orbital polarization,
so that  the system might actually  become a $d^1$ analog of $d^9$ cuprates. 
In this Letter, we examine this idea by means of LDA+DMFT.  

\begin{figure}[t]
\begin{center}
\includegraphics[width=8.0cm]{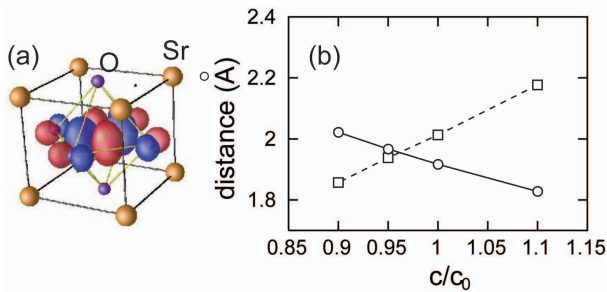}
\end{center}
\vspace{-.4cm}

\caption{
(a) Atomic configuration of Sr$_2$VO$_4$ together with the NMTO
Wannier function of the V $d_{xy}$-orbital.
The position of Sr and the apical oxygen are optimized 
by a plane-wave GGA calculation.
(b) GGA-optimized distance between O and V along the $c$ direction
(open squares) and in the $ab$-plane (open circles)
as a function of the $c$ axis elongation  $c/c_0$.
}
\label{Fig:fig1}
\end{figure}

{\em GGA optimization of crystal structure.} 
Let us now turn to our  actual calculations. First, 
we perform a GGA (generalized gradient approximation)
calculation with plane-wave basis, employing the
Tokyo Ab-initio Program Package (TAPP)\cite{Yamauchi96}.
Note that a plane-wave basis set has  advantages for
optimizing atomic configurations
under high pressure which are unknown experimentally.
We adopt the exchange-correlation functional introduced 
by Perdew {\it et al.}\cite{Perdew96} and ultra-soft 
pseudopotentials in a separable form\cite{Vanderbilt90,Laasonen93}.
The wave functions are expanded up to a cut-off energy 
of 36.0 Ry, and $8\times 8\times 8$ and $12\times 12\times 12$ 
$k$-point grids are used.
We assume that the system has the same $I4/mmm$ symmetry 
as at atmospheric pressure \cite{notetilt},
so that there are only two free parameters, i.e.,
the position of the apical oxygen and Sr. 
In Fig.\ \ref{Fig:fig1}(a), we show the atomic 
configuration of the octahedron around the V ion.

We  calculate the total energy 
as a function of the lattice constant $a$, fixing the ratio 
$c/a$ to its experimental value (=3.28)\cite{Nozaki}. 
We find that the energy minimum is at $a$=3.89\AA (not shown), 
in excellent agreement with experiment ($a$=3.84\AA). 
When the lattice constant $a$ is fixed to 
this optimized value, the Sr-V and O-V distances 
along the $c$-direction become 4.46\AA\ and 2.01\AA,
respectively, consistent with the experimental values
4.46\AA\ and 1.98\AA.

Next, we change the lattice constants to simulate
the effect of pressure.
Considering to apply high uniaxial pressure or, more realistically, to synthesize 
Sr$_2$VO$_4$ films on substrates with appropriate lattice constant,
 we change the lattice constant $c$ up to $\pm$10\%
of the experimental value $c_0=$12.76\AA.

In Fig.\ \ref{Fig:fig1}(b), we plot the GGA-optimized 
distance between 
O and V along the $c$-direction ($d_c$) and within the $ab$-plane
($d_{ab}$) as a function of $c/c_0$. 
Here we fix the volume of the unit cell ($V$) rather than $a$,
since the total energy at fixed $V$ is always lower 
than that of fixed $a$. While the V-O distance
 $d_c$ is longer than $d_{ab}$ at ambient pressure, 
 $d_c$ becomes shorter  than $d_{ab}$ for $c/c_0<0.95$. 

In fact, the ratio $d_c/d_{ab}$ 
determines the splitting of the three $t_{2g}$ orbitals.
If we press uniaxially along the $c$-direction,
the negatively charged oxygen
ions move towards the vanadium site. 
Hence the energy of the $d_{yz/zx}$-orbitals,
which point along the $c$-direction, is enhanced.
 At ambient pressure
the level splitting is small.
But
 given the 
fact that $d_c$ and $d_{ab}$ change considerably 
with $c/c_0$ in Fig.\ \ref{Fig:fig1}(b), we may 
expect that we can control the level splitting, and,
consequently, the 
orbital occupation by applying pressure.

\begin{figure}[t]
\includegraphics[width=8.5cm]{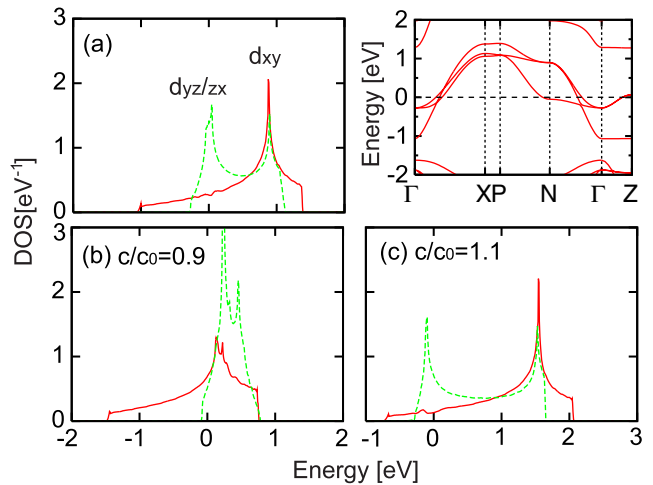}
\caption{Density of states of the $t_{2g}$ band for 
(a) atmospheric-pressure,
(b) uniaxial pressure in the $c$ direction ($c/c_0=0.9$), and
(c) uniaxial pressure in the $ab$ plane ($c/c_0=1.1$). 
Solid (dashed) line is for the $d_{xy}$-orbital 
($d_{yz/zx}$-orbitals).
Inset: LMTO  band structure for atmospheric-pressure.}
\label{Fig:fig2}
\end{figure}

{\em LDA+DMFT calculation.}
To examine this idea in the presence
of electron correlations, we perform
 LDA+DMFT calculations for
the atomic configurations obtained above.
To this end, we first carry out LDA band structure
calculations with the LMTO
(linearized muffin tin orbital) basis\cite{LMTO}. In the inset of
Fig.\ \ref{Fig:fig2}, we show the obtained band structure
for ambient pressure. Almost the same band structure 
is obtained by the GGA calculation with plane-wave basis.

Then, we extract the three $t_{2g}$ bands by the NMTO ($N$-th order muffin tin
orbital) downfolding\cite{NMTO},
using the generated LDA potential. 
As a typical example of the resulting 
NMTO Wannier functions,  we show
the $d_{xy}$-orbital 
of V 
in Fig.\ \ref{Fig:fig1}(a) for ambient pressure.  The density of states (DOS) of
the $t_{2g}$ band for $c/c_0=0.9,1.0$ and $1.1$ are shown 
in Fig.\ \ref{Fig:fig2}.
The NMTO band width of the $d_{xy}$-band 
is 2.46 eV at ambient pressure ($c=c_0$). 
This value is consistent with that of
the GGA calculation with plane-wave basis and that of
Pickett {\it et al.} who used the
full-potential linearized
augmented plane wave (LAPW) basis\cite{Pickett}.
Imai {\it et al.}\cite{Imai05} reported a smaller 
band width of $\sim$2.0 eV\cite{Comment}.

For the case of $c/c_0$=0.9(1.1), Fig.\ \ref{Fig:fig2}
shows that the center of gravity of the
$d_{yz/zx}$-bands is  clearly higher(lower) than that of the
$d_{xy}$-band.
Indeed, the crystal field splitting 
between the $d_{xy}$- and $d_{xz/yz}$-orbitals in the 
NMTO Hamiltonian is -382(+434) meV for $c/c_0$=0.9(1.1). 
Concerning the electron occupation  of the $d_{xy}$- and $d_{xz/yz}$-orbitals,
 90\%(20\%) of the $d$ electrons are accommodated
in the $d_{xy}$-band for $c/c_0$=0.9(1.1), in contrast to
ambient pressure where all three $t_{2g}$ bands are similarly occupied.


Next, we perform DMFT calculations for the three low-energy
 $t_{2g}$ bands, studying whether electronic correlations
result in a full  orbital polarization
 for $c/c_0$=0.9 and 1.1. To this end,
the DMFT effective impurity model is solved by the
standard Hirsch-Fye quantum Monte Carlo (QMC) method\cite{HF},
where the temperature is 0.1 eV with 100 imaginary time 
slices and the number of QMC sample is $\sim$ $2\times10^6$.
We employ the relation 
$U=U'+2J$ where
$U$, $U'$, $J$ are the intra-orbital Coulomb 
interaction, the inter-orbital Coulomb interaction and
the Hund coupling, respectively.

\begin{figure}[t]
\includegraphics[width=8.5cm]{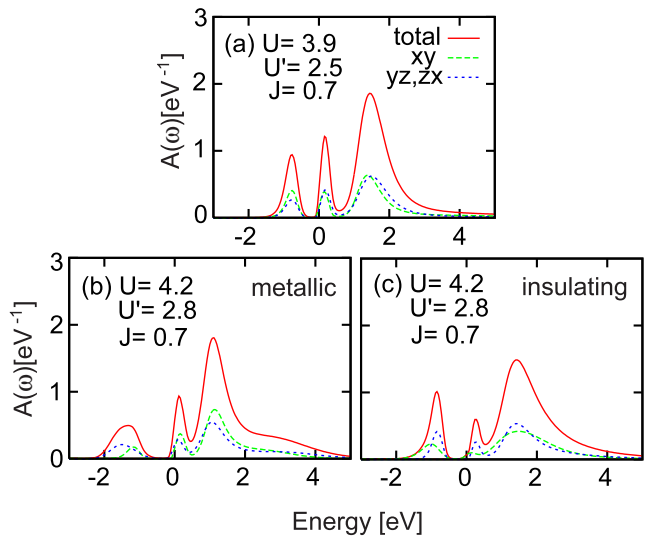}
\caption{LDA+DMFT spectral function for atmospheric-pressure, 
where dashed, dotted, and solid line are for
$d_{xy}$, $d_{xz/yz}$, and the total  Sr$_2$VO$_4$ spectrum, respectively. 
For  $U'\!=\!2.8$, two 
solutions [metallic (b) and insulating (c)] coexist.
\label{Fig:fig4}}
\end{figure}

We first calculate
the spectral function for ambient pressure 
with various interaction parameters $U'$,
fixing $J\!=\!0.7$ eV which is a reasonable value
for V ions. In Fig.\ \ref{Fig:fig4}, we show
the spectral function for $U'\!=\!2.5$ eV (a) and $2.8$ eV(b,c);
the Mott-Hubbard transition occurs 
around $U'\!=\!2.5\! \sim\! 2.8$ eV. 
Since the gap in (c)
is in accord with 
the main optical peak in experiment\cite{Matsuno05},
we  expect $U'\!\sim\!2.8$ eV for Sr$_2$VO$_4$.
The coexistence of this insulating solution (c)
with a metallic one (b)
indicates insulating Sr$_2$VO$_4$
to be close to a Mott-Hubbard 
transition.

Note that Imai {\it et al.}
\cite{Imai05} estimated smaller Coulomb interactions:
$U'\sim 1.3$ and $J\sim 0.65$ eV by the combination of
the constrained LDA and the GW method\cite{Imai05}.
However, as mentioned above, the band width of the $t_{2g}$-orbitals of
Ref.\ \cite{Imai05} is  20-25\%
smaller than that of the present study. 
If we normalize the interaction
parameters by the band width, the difference is not so big. 
On the other hand, Sekiyama {\it et al.} employed 
 $U'\!=\!3.55$ and $J\!=\!1.0\,$eV in their LDA+DMFT 
calculation for SrVO$_3$, reproducing the photoemission 
spectrum  of
SrVO$_3$\cite{Sekiyama}. 
These values
are  not so far from ours.

\begin{figure}[h]
\includegraphics[width=8.5cm]{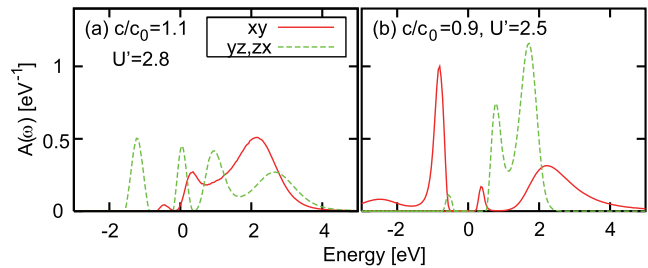}
\caption{Same as Fig.\ \ref{Fig:fig4} but for
(a) uniaxial pressure in the  $ab$ plane  ($c/c_0\!=\!1.1$) with $U'\!=\!2.8$ 
and (b) uniaxial pressure in the $c$ direction ($c/c_0\!=\!0.9$) 
with $U'\!=\!2.5$. 
}
\label{Fig:fig5}
\end{figure}

Let us now turn to the LDA+DMFT results for high pressure.
In Fig.\ \ref{Fig:fig5}(a), we plot the spectral function
for $c/c_0\!=\!1.1$, showing a metallic peak  at the Fermi
level. In contrast to ambient pressure, there is no  
coexisting insulating solution  at $c/c_0\!=\!1.1$, i.e.,
applying pressure in the $ab$ plane 
makes  Sr$_2$VO$_4$ metallic.
An important point is that the $d_{xy}$-orbital is
almost empty for $c/c_0\!=\!1.1$:
orbital polarization occurs. 
The system becomes a  quarter-filled 
2-band Hubbard model which is well known
to have a ferromagnetic ground 
state\cite{Momoi,Held98}.
Therefore, we expect ferromagnetic spin fluctuation 
to be dominant at low $T$ if pressure is applied in the $ab$ plane.

On the other hand, Fig.\ \ref{Fig:fig5}(b) shows the result for 
uniaxial pressure along the $c$ direction
($c/c_0=0.9$). We see that the spectrum 
is now clearly insulating, even for the smaller
value of  $U'\!=\!2.5$ for which
we have  a metal in  Fig.\ \ref{Fig:fig4}(a).
This is a surprising result: 
 an inverse Mott-Hubbard transition.
(Usually, applying pressure makes an
insulator metallic, not a metal insulating
as from  Fig.\ \ref{Fig:fig4}(a) to Fig.\ \ref{Fig:fig5}(b).)
What is the reason for this unusual behavior?
From
Fig.\ \ref{Fig:fig2}, we see that the LDA band width
does not change strongly from 2.26 eV at 
$c/c_0=1$ to  2.46 eV at $c/c_0=0.9$.
This small change of band width alone would indeed
indicate more metallic behavior--as usual.
But more important is that uniaxial pressure 
changes the crystal-field splitting:
The two $d_{yz/zx}$-orbitals become
unoccupied, and then  the large intra-orbital 
repulsion $U$ makes the remaining (single) $d_{xy}$-orbital 
Mott-insulating. This way, a $d^1$ analog of $d^9$ 
cuprates is realized.
Concerning the  magnetic properties, 
we expect an antiferromagnetic instability 
since the system becomes a half-filled 
single-band model.


{\em Ba$_2$VO$_4$.} 
While it is an interesting possibility to change
the electronic properties by controlling the lattice constant,
it might be difficult to change its value up to $\pm$ 10\%.
For exapmle, in order to grow Sr$_2$VO$_4$ thin films with c/c$_0$ = 0.9, 
we need considerably large lattice mismatch between a substrate 
and bulk Sr$_2$VO$_4$, while an excessive mismatch often results in 
lattice relaxation.

Thus, lastly, we consider the possibility of chemical
pressure by substituting Sr by Ba. Since the ion radius of Ba is larger 
than that of Sr, 
the crystal is expected to be expanded. 
The important point here is that the V-O distance
in the $ab$ plane and that along the $c$ axis 
will increase differently. Namely, while the increase of
 $a$ directly affects 
the V-O distance in the $ab$ plane (the latter is
exactly half of the former), $c$ and the V-O distance 
along the $c$ axis are independent parameters.

First, we perform GGA calculation with structure optimization 
for Ba$_2$VO$_4$ by changing $a$ and $c$. 
The energy minimum is at $a$=4.04 \AA\ and $c/a=3.36$,
the  V-O distances are  2.02  \AA\ in the $ab$ plane 
(compared to 1.92  \AA\ for  Sr$_2$VO$_4$) and
2.01 \AA\ along the $c$ direction (2.01 \AA\ for  Sr$_2$VO$_4$).
This means that the situation for  Ba$_2$VO$_4$ is similar to that
of  Sr$_2$VO$_4$ 
with $c/c_0=0.95$, i.e., if 
uniaxial pressure is already applied.

\begin{figure}[tb]
\includegraphics[width=8.5cm]{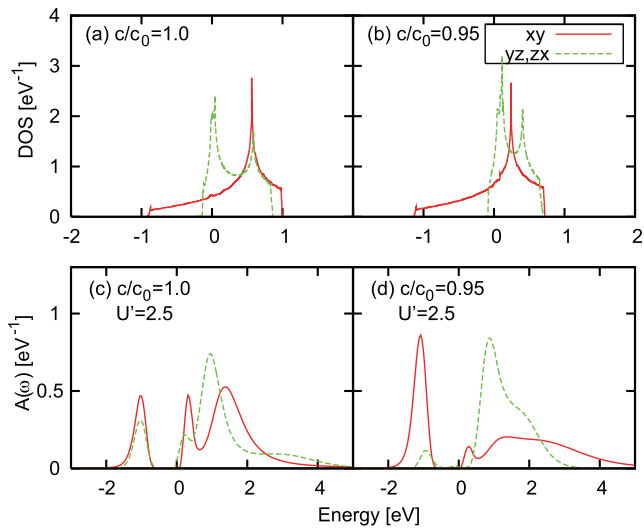}
\caption{
Density of states of the $t_{2g}$ band of Ba$_2$VO$_4$
for (a) atmospheric-pressure and (b) uniaxial pressure in 
the $c$ direction ($c/c_0\!=\!0.95$). Figs.
(c) and (d) are LDA+DMFT spectral function 
for (a) and (b) respectively, with $U'\!=\!2.5$.
}
\label{Fig:fig6}
\end{figure}

Next, we consider the compression $c/c_0=0.95$ for Ba$_2$VO$_4$.
The orbital polarization becomes larger, {\it i.e.},
the electron density is 0.73 for the $d_{xy}$ band and 0.14 
for the $d_{yz/zx}$ band. Thus it is interesting to 
proceed with DMFT calculations. In Fig. \ref{Fig:fig6}, we plot 
the resulting LDA+DMFT spectral function for ambient pressure and
uniaxial pressure in the $c$ direction ($c/c_0\!=\!0.95$)
with $U'=2.5$, along with the density of states by LDA.
We can see that a large orbital polarization is realized
even for $c/c_0\!=\!0.95$. Since 
the relative hopping and interaction  parameters for the
 $d_{xy}$ band are very similar
to the cuprates, a $d^1$ analog of $d^9$ 
cuprates can be realized.
Experimentally, the uniaxial pressure is best realized by
growing  Ba$_2$VO$_4$ on a substrate with a larger in-plane 
lattice constant 4.1-4.2 \AA.

%

We thank J. Akimitsu, O.\ K.\ Andersen, H. Aoki,
  H.-U. Habermeier, M. Imada, Y. Imai, 
K. Kawashima, G. Khaliullin, I. Solovyev 
and Y.-F. Yang
for fruitful discussions
and the  Alexander von Humboldt foundation (RA) and the 
Emmy Noether program of the Deutsche Forschungsgemeinschaft (KH)
for financial support.
Numerical calculations were done at the Supercomputer Center, 
Institute for Solid State Physics, University of Tokyo.


\begin{references}
\bibitem{Bednorz86}
J.\ G.\ Bednorz and K.\ A.\ M\"uller, Z.\ Phys B {\bf 64}, 189 (1986).
\bibitem{Kohn}P.\ Hohenberg and W.\ Kohn, Phys. Rev. {\bf 136}, B864 (1964),
W.\ Kohn and L.\ J.\ Sham,  Phys. Rev. {\bf 140}, A1133 (1965).
\bibitem{beyondLDA} See, {\it e.g.}, {\it Strong Coulomb 
Correlations in Electronic Structure Calculations: 
Beyond the Local Density Approximation}, Ed. by V.\ I.\ Anisimov,
G \& B Science Pub. (1999).
\bibitem{Anisimov}
V.\ I.\ Anisimov {\it et al.}, J. Phys. Condens. Matter
{\bf 9}, 7359 (1997), A.\ I.\ Lichtenstein and M.\ I.\ Katsnelson,
Phys. Rev. B {\bf 57}, 6884 (1998).
\bibitem{Held}
K.\ Held {\it et al.}, Psi-k Newsletter {\bf 56}, 65 (2003) 
[phys. stat. sol.  (b) {\bf 243}, 2599 (2006)];
cond-mat/0511293; G. Kotliar {\em et al.}, 
Rev. Mod. Phys. {\bf 78}, 865 (2006).
\bibitem{Maeno03}
A.\ P.\ Mackenzie and Y.\ Maeno, Rev. Mod. Phys. {\bf 75} 657 (2003), 
and references therein.
\bibitem{Matsuno05}
J.\ Matsuno, Y.\ Okimoto, M.\ Kawasaki and Y.\ Tokura,
Appl. Phys. Lett. {\bf 82}, 194 (2003),
Phys. Rev. Lett. {\bf 95}, 176404 (2005).
\bibitem{Pickett}W.\ E.\ Pickett {\it et al.}, 
Physica C 162-164 1433 (1989).
\bibitem{Nozaki}
M.\ Rey {\it et al.}, J. Solid State Chem. {\bf 86} 101 (1990),
A.\ Nozaki {\it et al.}, Phys. Rev. B. {\bf 43} 181 (1991),
V.\ Ginnakopoulou, P.\ Odier, J.\ M.\ Bassat and J.\ P.\ Loup,
Solid State Commun. {\bf 93} 579 (1995).
\bibitem{Imai05}
Y.\ Imai, I.\ Solovyev, and M.\ Imada,
Phys. Rev. Lett. {\bf 95} 176405 (2005).
Y.\ Imai and M.\ Imada, J. Phys. Soc. Jpn. {\bf 75}, 094713 (2006).
\bibitem{Weng05}H.\ Weng, K.\ Kawazoe, X.\ Wan and J.\ Dong,
Phys. Rev. B {\bf 74}, 205112 (2006).
\bibitem{Yamauchi96}
J.\ Yamauchi, M.\ Tsukada, S.\ Watanabe,
and O.\ Sugino, Phys. Rev. B {\bf 54}, 5586 (1996).
\bibitem{Perdew96}
J.\ P.\ Perdew, K.\ Burke, and Y.\ Wang, Phys. Rev. B
{\bf 54}, 16533 (1996).
\bibitem{Vanderbilt90}
D.\ Vanderbilt, Phys. Rev. B {\bf 41}, 7892 (1990).
\bibitem{Laasonen93}K.\ Laasonen, A.\ Pasquarello, R.\ Car, C.\ Lee,
and D.\ Vanderbilt, Phys. Rev. B {\bf 47}, 10142 (1993).
\bibitem{notetilt} 
Because of the large radius of the Sr ion, a significant
tilting
of the oxygen octahedra is not to be expected. 
\bibitem{LMTO} O.\ K.\ Andersen, Phys. Rev. B {\bf 12}, 3060 (1975),
O.\ K.\ Andersen and O.\ Jepsen, Phys. Rev. Lett {\bf 53}, 2571 (1984).
\bibitem{NMTO} O.\ K.\ Andersen and T.\ Saha-Dasgupta, 
Phys. Rev. B {\bf 62}, R16219 (2000) and references therein.
\bibitem{Comment}The difference can be explained by 
different support points for linearization/$N$-ization.
(I.\ Solovyev, private communication.)
\bibitem{HF} J.\ E.\ Hirsch and R.\ M.\ Fye, Phys. Rev. Lett.
{\bf 56} 2521 (1986).
\bibitem{Sekiyama} A.\ Sekiyama {\it et al.}
Phys. Rev. Lett. {\bf 93}, 156402 (2004).
\bibitem{Momoi}  T.\ Momoi and K.\ Kubo, 
Phys. Rev. B {\bf 58} R567 (1998). 
\bibitem{Held98} K.\ Held and D.\ Vollhardt, 
Eur. Phys. J. B {\bf 5} 473 (1998).
\end{references}
\end{document}